\documentclass[journal=amlccd,manuscript=letter,layout=traditional]{achemso}

\usepackage[version=3]{mhchem} 
\usepackage{subfigure}
\usepackage{color}

\usepackage[T1]{fontenc}  

\author{Thomas C. O'Connor}
\affiliation{Department of Physics and Astronomy, Johns Hopkins University, Baltimore, MD 21218 USA}
\email{toconnor@jhu.edu}
\author{Mark O. Robbins}
\email{Mark.O.Robbins@jhu.edu}
\affiliation{Department of Physics and Astronomy, Johns Hopkins University, Baltimore, MD 21218 USA}
\title[\texttt{achemso} demonstration]
{Chain Ends and the Ultimate Strength of Polyethylene Fibers}

\begin{document}
\begin{abstract}
We use large scale molecular dynamics (MD) simulations to determine the tensile yield mechanism of orthorhombic polyethylene (PE) crystals with finite chains spanning $10^2$--$10^4$ carbons in length. We find the yield stress $\sigma_y$ saturates for long chains at 6.3 GPa, agreeing well with experiments. We show chains do not break but always yield by slip, after nucleation of 1D dislocations at chain ends. Dislocations are accurately described by a Frenkel-Kontorova model parametrized by the mechanical properties of an ideal crystal. We compute a dislocation core size $\xi\approx25$\,\AA\ and determine the high and low strain rate limits of $\sigma_y$. Our results suggest characterizing the 1D dislocations of polymer crystals as an efficient method for numerically predicting the ultimate tensile strength of aligned fibers.
\end{abstract}


Ultra-high-molecular-weight polyethylene (UHMWPE) fibers are highly ordered materials with up to 95\% crystallinity and near perfect chain alignment. Measurements of elastic moduli range between 150-300GPa, approaching metallic strengths but at a fraction of the weight and cost.\cite{Crist1995,kelly1986,Tashiro1978,Werff1991} Applications have rapidly grown to include cables, fabrics, composite armors and glasses.\cite{Marissen2011}

\begin{figure}[htb]
\includegraphics[width=0.45\textwidth]{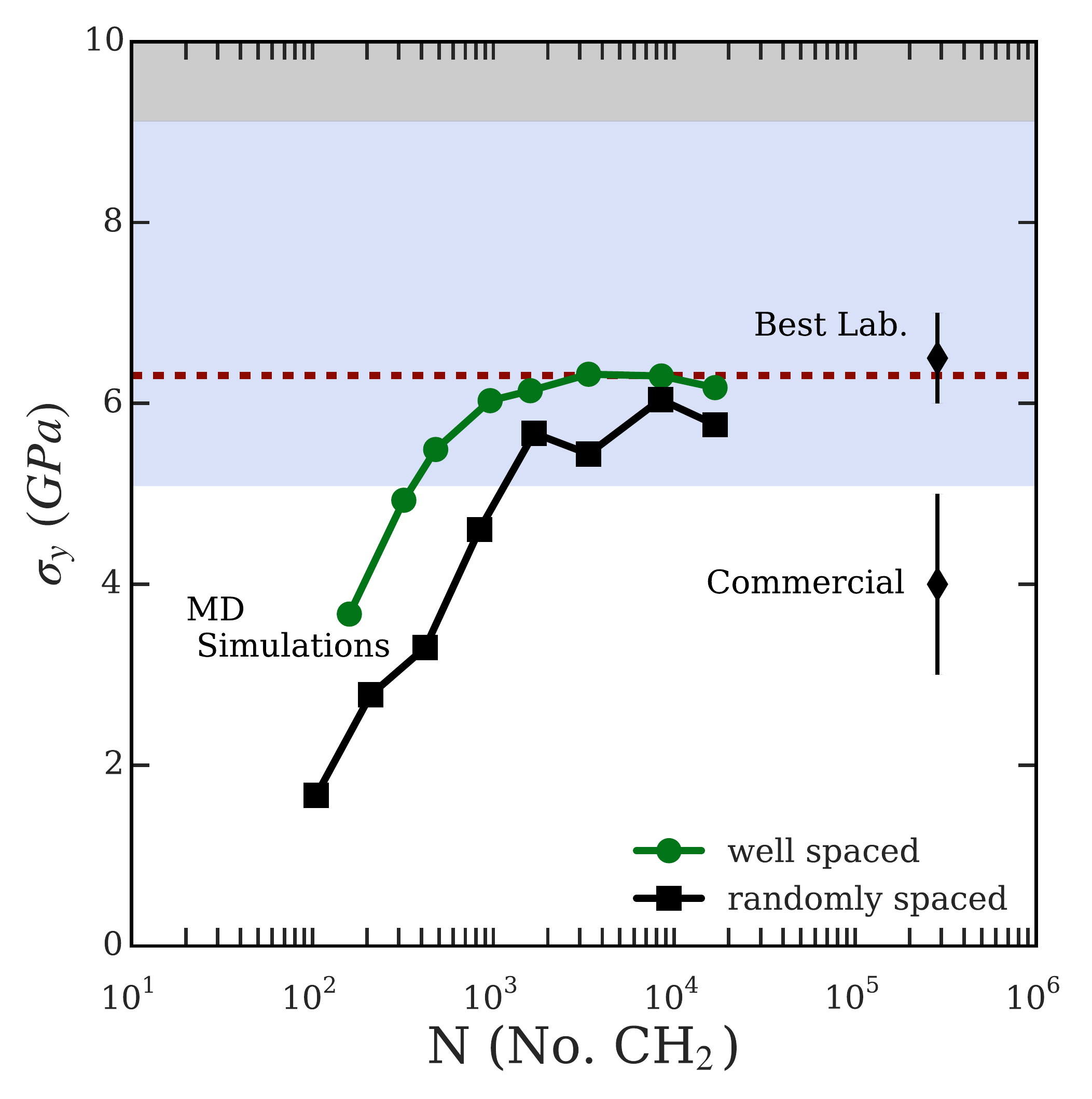}
\caption{Yield stress $\sigma_y$ vs number of carbons per chain $N$ at 300K for simulations with well-spaced (green circles) and random (black squares) chain end distributions compared to experimental strengths for commercial\cite{Hudspeth2012,Hudspeth2015} and lab\cite{Hoogsteen1988d,Hoogsteen1988e,Gorshkova1987} fibers (black diamonds). Experimental chain lengths range from $N=10^5-10^7$. The dashed red line indicates the stress where an isolated chain end defect slips by thermal activation of a dislocation at 300K. Shaded regions show FK model predictions for chain end metastability (blue) and instability (gray) against slip by dislocation nucleation.}
\label{fig:yieldstress}
\end{figure}

The ordered fiber state is usually obtained by plastically straining (drawing) a UHMWPE gel. The drawing process aligns chains and facilitates the formation of large ordered crystalline domains with a uniform orientation, leading to a high strength in the direction of alignment.\cite{McDaniel2015} Advances in processing have led to commercial PE fibers that yield at strains $\epsilon_y\sim 0.02-0.04$ with tensile strengths of about $\sigma_y\sim 3-4$GPa.\cite{Hudspeth2012,Hudspeth2015} Strengths of 6-7GPa\cite{Hoogsteen1988d,Hoogsteen1988e,Gorshkova1987} have been obtained in the laboratory by increasing chain alignment and crystallinity (black diamonds in Figure \ref{fig:yieldstress}). While impressive, first-principal calculations predict that carbon bonds along the backbone would allow strengths four times higher ($\epsilon_y\sim 0.10$ and $\sigma_y\sim 20-40$GPa) before breaking. \cite{Boudreaux1973,Suhai1986} Classifying the defects and yield mechanisms that cause the reduction in strength remains a topic of active interest.

Real PE fibers are made of finite-length chains and thus have chain end defects (see appended graphic). Chain ends allow the fiber to yield by chains slipping past each other. Such chain slip is mediated by intermolecular van der Waals bonds ($\sim 10^1$ meV), which are much weaker than the intramolecular C-C bonds ($\sim4$eV). Mesoscale investigations have shown that atomic scale studies are needed to accurately quantify these weak van der Waals forces and determine if fibers yield primarily by chain slip, bond breaking, or some combination of the two.\cite{Termonia1985,Smook1984,Dijkstra1989b}

In this letter we use large scale MD simulations to determine the tensile strength of crystalline PE, which is believed to provide an upper bound for PE fibers. We find that infinite chains fail through chain scission at $\sim 20$GPa, which is consistent with previous estimates. In contrast, introducing chain ends leads to failure by chain slip. Simulated crystals with finite chain lengths ranging from $10^2-10^4$ carbon atoms yield by slip at chain ends and give a limiting tensile yield stress of 6.3GPa that is consistent with the best laboratory fibers. The yield stress, its dependence on chain length, and its high and low rate limits can be understood using a simple Frenkel-Kontorova model with parameters extracted from simulations of the ideal crystal.

We model PE with AIREBO-M, a reactive bond-order potential for hydrocarbons.\cite{OConnor2015} Its van der Waals interactions have been optimized for alkanes and give  accurate PE crystal phase behavior.\cite{OConnor2015} The stable phase of PE at T=300K is orthorhombic, described by the three lattice constants $\left(a,b,c\right)$. The constant $c=2.54$\AA\ is the $\text{C}_2\text{H}_4$ monomer spacing which points along the chain axis. We align the three lattice directions with the $\left(x,y,z\right)$ axes of our periodic simulation cell. There are 80 chains in the xy-plane of the cell. The period along the z-axis is N/2 monomer spacings, corresponding to $N=105-16800$ $C$ atoms. The box dimensions are about  $\left(L_x,L_y,L_z\right)\approx\left(36.77,39.53,2.54\times N/2\right)$\,\AA.

We make finite-length chains by deleting one monomer ($\text{C}_2\text{H}_4$) from each chain's backbone, breaking the chain and creating two chain ends. Newly exposed chain ends are sp2 hybridized so we add an extra hydrogen atom to each to saturate the missing bond. This yields periodic crystals with an axial length of N carbons, composed of finite chains that are N-2 carbons in length. The spatial distribution of chain end defects can significantly impact crystal strength.\cite{Kausch1979} We address this by distributing defects in two ways. In the first set of systems, locations are selected so that defects on nearest and next nearest neighbor chains have large axial separation ($>\frac{1}{20}Nc$). In the second set, defects have a random axial distribution.

We equilibrate all systems to T=300K and 0 pressure using the LAMMPS software package with a Nos\'{e}-Hoover NPT ensemble and a time-step between 0.5-1.0 fs.\cite{Plimpton1995} Once equilibrated, we load each crystal in uniaxial tension by applying a constant engineering strain rate along the z-axis up to 5\% strain, while allowing the stress in x and y to relax to 0GPa. Strain rates $\dot\gamma$ range from $\sim9\times10^{6}$ -- $10^9\ s^{-1}$. Rates are selected to fix  $\dot\gamma N$ so that strains can be accommodated by chain ends moving at a velocity that is independent of the chain length N. 

\begin{figure}[htb]
\includegraphics[width=0.45\textwidth]{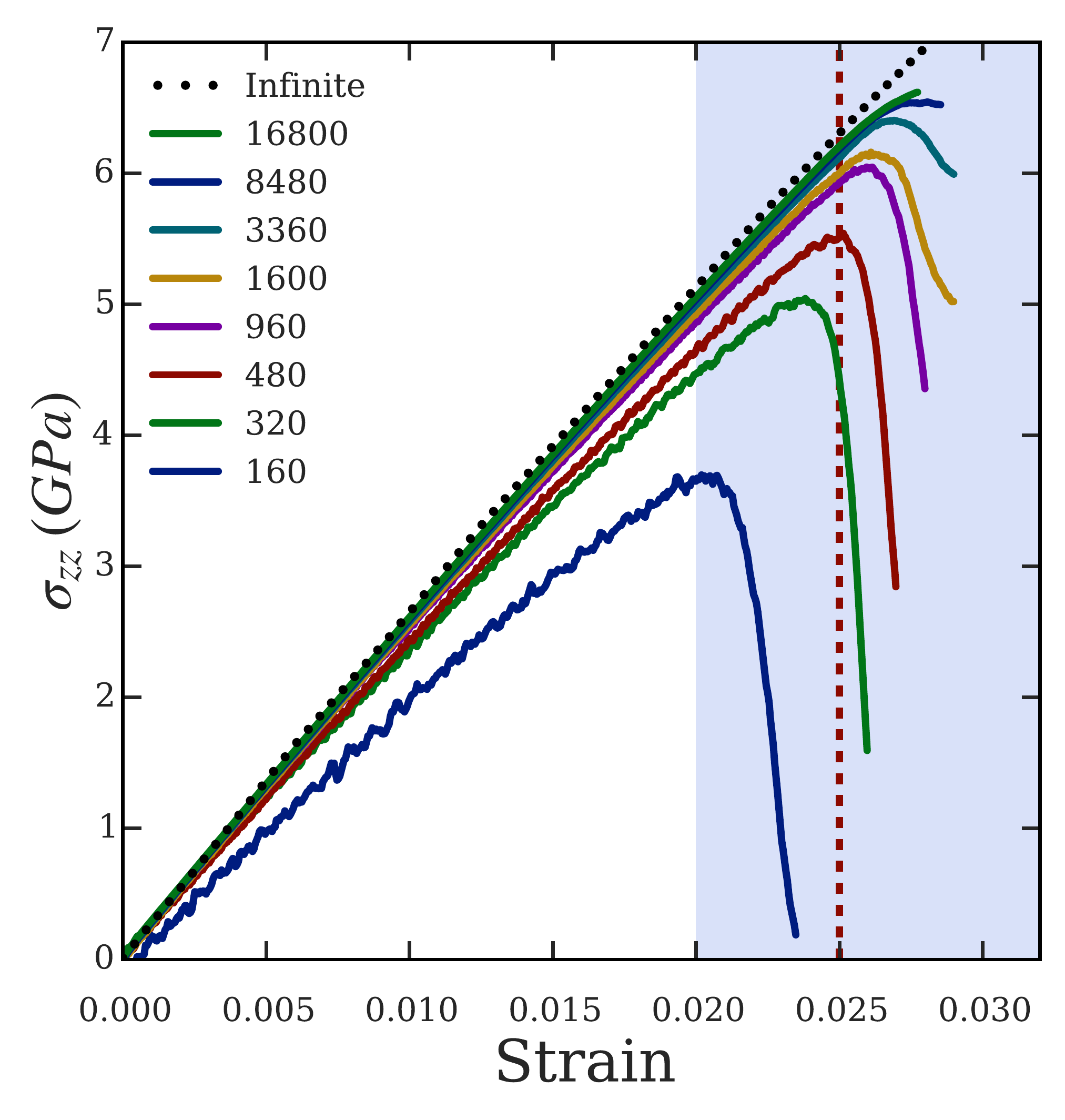}
\caption{$\sigma$ vs. $\epsilon$ curves for uniaxial tension tests of crystals with well-spaced defects at T=300K (solid lines) and a defect-free periodic crystal (black dotted line). The mechanical response saturates as $N$ increases with $E\to258$GPa, $\sigma_y \to6.3$GPa, and $\epsilon_y \to0.025$ (vertical dashed line). The defect free crystal yields by scission at much higher stress $\sim20$ GPa.}
\label{fig:moduli}
\end{figure}

Stress strain curves for the uniaxial tension tests are plotted in Figure \ref{fig:moduli}. All systems show an initial linear response. For $N>1000$ the modulus  $E$ saturates at $\sim258$GPa, which is comparable to experiments and theory for ideal crystals.\cite{Werff1991,Suhai1986} As strain increases, the tangent modulus, $d\sigma/d\epsilon$, shows a sudden drop indicating the initiation of failure. We take this as the yield stress ($\sigma_y$) since each crystal fails if the stress is maintained at this value. Peak stresses are sometimes  slightly higher ($\sim0.3$GPa) but are sensitive to the high strain rates in our simulations, implying post yield effects.

For infinite chains, chain scission is observed at yield and the yield stress of $\sim20$GPa is consistent with estimates based on carbon bond strength.\cite{Boudreaux1973,kelly1986} For all finite chains the yield stress is much lower, no bond scission is observed, and yield is associated with chain slip. Figure \ref{fig:yieldstress} shows $\sigma_y$ as a function of N for crystals with different arrangements of chain ends. Maximizing the separation between defects on nearby chains leads to $\sigma_y=6.3$GPa with a saturation in strength at about $N=1000$. Breaking bonds at random gives a slightly lower $\sigma_y$ and slower saturation. This reduction in strength can be associated with the minimum separation between chain end defects. In general we find saturation at a stress comparable to the best laboratory crystals, at chain lengths much less than the $N \sim 10^5-10^7$ in experiments.\cite{stein1988ultrahigh,Hoogsteen1988d,Hoogsteen1988e} Detailed analysis of a range of configurations shows that the key limitation is the minimum separation of chain end defects. Lab fibers have much longer chains than our systems and are expected to be in this saturated limit.\cite{Crist1995}

\begin{figure}[htb]
\includegraphics[width=0.45\textwidth]{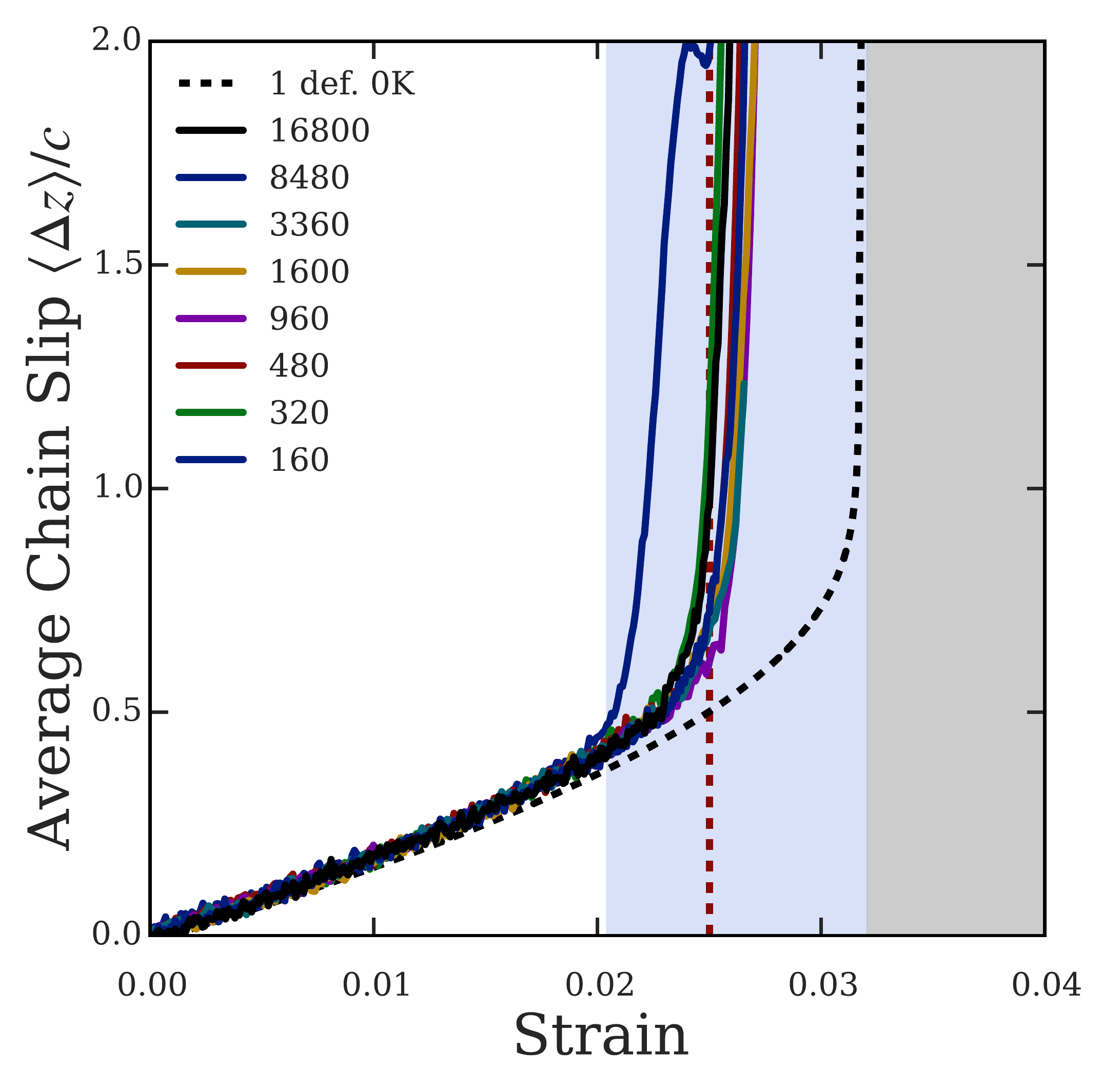}
\caption{Average chain end seperation $\left<\Delta z\right>$ vs. strain at 
T=300K for systems with well-spaced defects (solid lines). The dashed black line is $\Delta z$ for an isolated defect in a perfect crystal at T=0K. The dashed red line indicates $\epsilon=0.025$, where an isolated defect at T=300K slips by thermal activation of dislocations. At small strains, all systems follow the T=0K curve of the isolated defect before slipping thermally. For large $N$, all systems slip near $\epsilon=0.025$.}
\label{fig:ends}
\end{figure}

We analyze the detailed changes in the molecular configurations at yield and find the yield mechanism to be slip at chain ends. A simple way of quantifying this is to measure the mean separation between chain end pairs $\left<\Delta z\right>$. In figure \ref{fig:ends} we plot $\left<\Delta z\right>$ normalized by the monomer spacing $c$ as a function of strain for systems with well-spaced defects (solid lines). At small strains, all systems show the same gradual rise in $\left<\Delta z\right>$, which represents the elastic response in the confining potential of surrounding chains. There is then a rapid rise, indicating yield by chain slip. Here we define the yield strain ($\epsilon_y$) for chain slip as the strain when $\left<\Delta z\right> > c=2.54$\,\AA\ - i.e., when the majority of chain ends have slipped by 1 monomer spacing. Once $N\geq320$, $\epsilon_y$ reaches a limiting value of $\sim0.025$, which is consistent with our analysis of the stress-strain curves (Figure \ref{fig:moduli}). 

Figure \ref{fig:ends} suggests $\epsilon_y$ becomes independent of $N$ when chain end defects are far apart and their interactions are weak. Since we have kept $\dot\gamma N$ constant, our well-spaced simulations with $N>320$ are at strain rates between $9\times10^6-2\times10^8\,s^{-1}$, but yield at nearly the same $\epsilon_y$. Additional simulations for the longest ($N=16800$) chains show that increasing $\dot\gamma$ from $9\times10^6s^{-1}$ to $5\times10^9s^{-1}$ only increases $\epsilon_y$ to 0.028, corresponding to $\sigma_y\sim7$GPa. Thus we expect little strength enhancement will be possible with experimentally accessible strain rates. 

We next investigate what sets the limiting value of $\epsilon_y$. In order to probe this limit, we seek a system that eliminates the interaction of chain ends on different chains. By repeating our previous analysis for a crystal with sufficient axial length (N = 480), but with only one of the 80 chains broken, we measure the slip and yield properties of a single finite chain in an otherwise perfect crystal. This system lets us study the stability and slip of a single chain end defect in the limit of "infinite" defect separation that is appropriate for laboratory chain lengths.

We first investigate the stability of the single defect in the absence of thermal fluctuations (T=0K), by performing a series of energy minimizations on the single defect crystal with the FIRE minimizer\cite{Bitzek2006} for $\epsilon\leq0.05$. The resulting curve of $\Delta z$ vs $\epsilon$ for the single defect is plotted in Figure \ref{fig:ends} as a black dashed line. We see that $\Delta z$ grows gradually with $\epsilon$ until the isolated defect becomes unstable to slip around $\epsilon\sim0.032\%$ and $\sigma\sim9.3$GPa. This stress and strain sets the upper limit of stability of the chain end defect. Past this strain, chain ends slip spontaneously, even in the absence of thermal fluctuations.

A finite chain can also be metastable and slip by thermal activation at finite temperature. To study thermal activation at 300K we held the system at a fixed strain and monitored $\Delta z$ to see if chains slipped in 1.0ns. Increasing strain in steps of 0.001 mimics strain rates of $10^6\,s^{-1}$, compared to $\dot\gamma$ up to $10^9s^{-1}$ in direct simulations. All configurations of the single defect with strain $\epsilon\geq0.025$ slip by thermal activation within 1.0 ns. This gives an upper bound on the stress and strain for thermally activated slip of $\sigma=6.3$ GPa and $\epsilon=0.025$ respectively. Plotting these values as red dashed lines in Figures \ref{fig:moduli}, \ref{fig:yieldstress}, and \ref{fig:ends}, we see the slip of a single defect at 300K effectively predicts $\epsilon_y$ and $\sigma_y$ for our PE crystals.

In order to understand the nature of chain slip, we must examine how stress is transferred between chains in the PE crystal. While an infinite chain can be loaded in tension directly along its backbone bonds, a finite chain can not. Instead the periodic potential produced by adjacent chains favors stretching to remain in registry, but at the expense of elastic energy. This competition between the elastic stiffness of a chain and shear stresses imposed by a periodic potential is the essence of the well-studied Frenkel-Kontorova (FK) model.\cite{frenkel1938,Braun2004}

The FK model describes the dynamics of a chain of masses connected by springs of stiffness $k$ resting on a periodic potential "substrate" with a period $\lambda$ that can be different from the equilibrium length of the springs, $c$. In our case, $c$ equals the equilibrium monomer spacing $c=2.54$\AA, but the period of the potential created by neighboring chains grows with strain as $\lambda=c\left(1+\epsilon\right)$. As strain increases, the peak force $\tau$ from the periodic potential is insufficient to balance the elastic tension of the stretching bonds, and the chain ends depin by nucleating 1D dislocations.  

Frank and van der Merwe were the first to consider 1D dislocations for fixed $k$ and mismatch $\lambda/c$.\cite{Frank1949} A large group of researchers has extended their ideas to describe a multitude of condensed matter systems.\cite{Braun2004} This includes many who have invoked the ideas of Frank and van der Merwe to describe the dynamics of PE and other polymer rystals.\cite{Kausch1973,Manevitch1997,Savin1998,Balabaev2001,Zubova2001a,Milner2011,Zubova2012,Hammad2015} None have considered the actual geometry of chain ends in the orthorhombic PE crystal under tension, instead they typically apply FK theory to simple bead-spring or united-atom potentials.\cite{Kausch1973,Manevitch1997,Savin1998,Balabaev2001,Zubova2001a,Hammad2015} We find that explicit hydrogens are essential to the shear modulus and interchain forces within the PE crystal. Shear forces are underestimated by a factor of $2-3$ in united-atom models. More detailed applications of FK theory to all-atom PE crystals consider short chains ($N<1000$) in unstressed crystals, and do not address yield.\cite{Milner2011,Zubova2012}

We find that an FK model, parametrized from all-atom simulations of the orthorhombic PE crystal, describes the tensile yield in our detailed finite-chain simulations as chain slip by dislocation nucleation at chain ends. The properties of the dislocations allow us to understand trends in yield with $N$ and place bounds on $\sigma_y$ that agree well with our simulations and experiments.

Only two parameters are needed to parameterize the FK model, $k$ and $\tau$. We determine a per monomer ($\text{C}_2\text{H}_4$) stiffness $k = 11.4\,eV/\text{\AA}^2$ from a uniaxial tension simulation of a defect free crystal. This is consistent with the experimental Young's modulus. We calculate the peak force per monomer of the periodic potential $\tau=46.72\,meV/\text{\AA}$ by displacing a single chain along the chain axis while measuring the force exerted on it by the surrounding crystal. The chain can relax in the transverse directions, but the surrounding crystal is held rigid by stiff harmonic constraints. This corresponds to a T=0K measurement of $\tau$. We find that $k$ and $\tau$ change little as T increases to room temperture.

\begin{figure}[htb]
\includegraphics[width=0.45\textwidth]{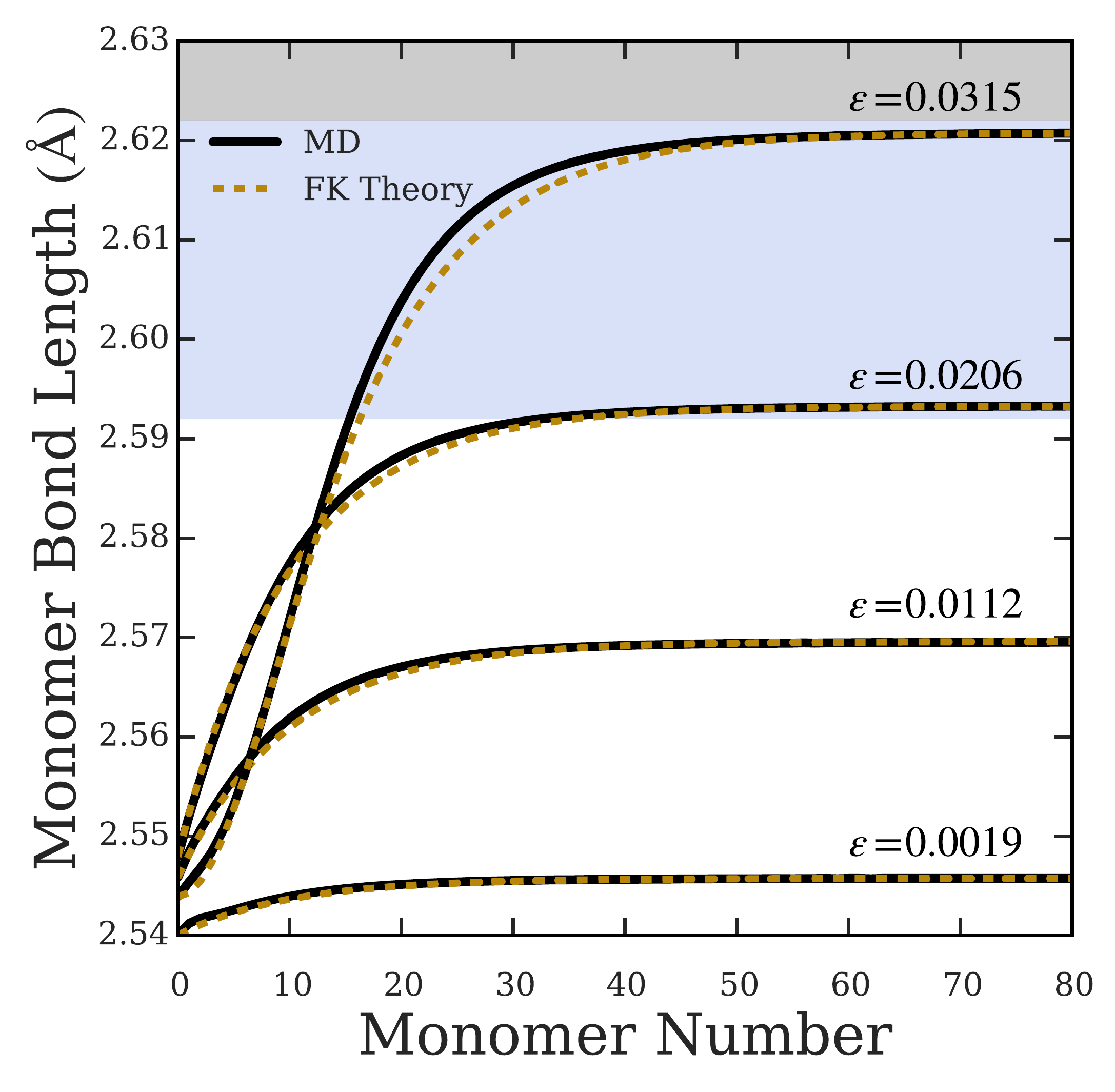}
\caption{Bond lengths as a function of distance from chain end from simulations (solid black) and theory (dashed lines) for the indicated strains. Interior bonds stretch uniformly with the crystal, while bonds near the end relax to the unstressed length.}
\label{fig:bonds}
\end{figure}

FK theory\cite{Frank1949,Braun2004} predicts a characteristic length over which the chain and periodic potential remain in registry called the dislocation core size  
$$\xi=\left(\frac{k c^3}{2\pi\tau}\right)^{1/2}=25.24\,\text{\AA}\approx 10\,c.$$ Tensile stress builds from 0 at the chain ends to the bulk value over a length $\sim\xi$. Over this scale, bond lengths must change from the relaxed length $c$ to the uniformly stretched length $c\left(1+\epsilon\right)$ in the chain center. Figure \ref{fig:bonds} compares numerical results and anlytical predictions for the change in bond length near chain ends. For small strains, $\epsilon\leq0.0206$, bond lengths follow the predicted simple exponential form $\sim exp(-x/\xi)$. At strains just before slip, the analytic prediction  takes the form of a dislocation $\sim exp(-x/\xi)/\left[1+exp(-2 x/\xi)\right]$. This captures our numerical result for $\epsilon=0.0315$ in Figure \ref{fig:bonds}. For larger strains, the chain end slips by a period or more, creating a dislocation that moves away from the end and leads to yield. 

The core size also sets the axial length-scale over which chain end defects on neighboring chains interact.\cite{Braun2004} As shown above in Figure \ref{fig:yieldstress}, $\sigma_y$ saturates at the strength of the isolated defect when $N\geq1600$, corresponding to a minimum separation of $80$ monomers ($\sim8\xi$) between defects on neighboring chains.

In the presence of chain ends, we can also derive the strains at which an end becomes metastable ($\epsilon_m$) and unstable ($\epsilon_u$) to dislocation formation (i.e. chain slip). The result, originally calculated by Frank and van der Merwe,\cite{Frank1949} gives: $\epsilon_m = \frac{2}{\pi^2}\left(\frac{c}{\xi}\right)=0.020,\quad \epsilon_u=\frac{1}{\pi}\left(\frac{c}{\xi}\right)=0.032.$
Using these values we have shaded regions in all Figures corresponding to  the metastability (blue) and instability (gray) of an isolated chain end to dislocation nucleation. The boundary between blue and gray regions represents an upper bound for yield: $\sigma_y \leq E\frac{c}{\pi\xi}=9.3\,\text{GPa,}$ that can only be reached in the limit of zero temperature or extremely high strain rate. It shows excellent agreement with our T=0K analysis plotted in Figures \ref{fig:ends} \& \ref{fig:bonds}. The lower boundary of the blue region corresponds to the low rate limit of $\epsilon_y$ and a lower bound for yield: $\sigma_y \geq E\frac{2\,c}{\pi^2\xi}=5.26\,\text{GPa,}$ where dislocations become energetically favorable, and creep-like yield can occur by thermally activated dislocation nucleation. Experimental values for PE crystals should fall somewhere between these bounds determined by rate and temperature effects. As noted above, we found increasing $\dot\gamma$ from $9\times10^6$ to $5\times10^9s^{-1}$ changes $\sigma_y$ less than 1GPa at room temperature. Thus experimental strengths are unlikely to exceed 7GPa.

In summary, all atom simulations of tensile failure show that the limiting strength of polyethylene fibers is determined by slip of chains near free ends. The yield stress grows with increasing chain length and saturates at values comparable to the best experimental fibers, 6.3GPa, and far below the stress required for chain scission. Little rate sensitivity ($<1$GPa) is seen for rates from $10^6$ to $10^9 s^{-1}$. The strength saturates when the separation between nearby chain ends is much larger than the dislocation size $\xi$ characterizing the length for transfer of tensile stress between adjacent chains. A simple Frenkel-Kontorova model parameterized by studies of small crystals captures the essential features of full simulations. The model provides analytic predictions for $\xi$, bond lengths, and the strains and stresses where the system is metastable and instable against chain slip. These can be related to bounds on the yield stress at low temperature and high and low rates. The approach is readily extended to other polymer fibers.

\begin{acknowledgement}
This research was performed within the Center for Materials in Extreme Dynamic Environments (CMEDE) under the Hopkins Extreme Materials Institute at Johns Hopkins University. Financial support was provided by the Army Research Laboratory under the MEDE Collaborative Research Alliance, through grant W911NF-12-2-0022.
\end{acknowledgement}


\bibliographystyle{achemso}
\providecommand{\latin}[1]{#1}
\providecommand*\mcitethebibliography{\thebibliography}
\csname @ifundefined\endcsname{endmcitethebibliography}
  {\let\endmcitethebibliography\endthebibliography}{}

\begin{figure}[pb]
\includegraphics[width=\textwidth]{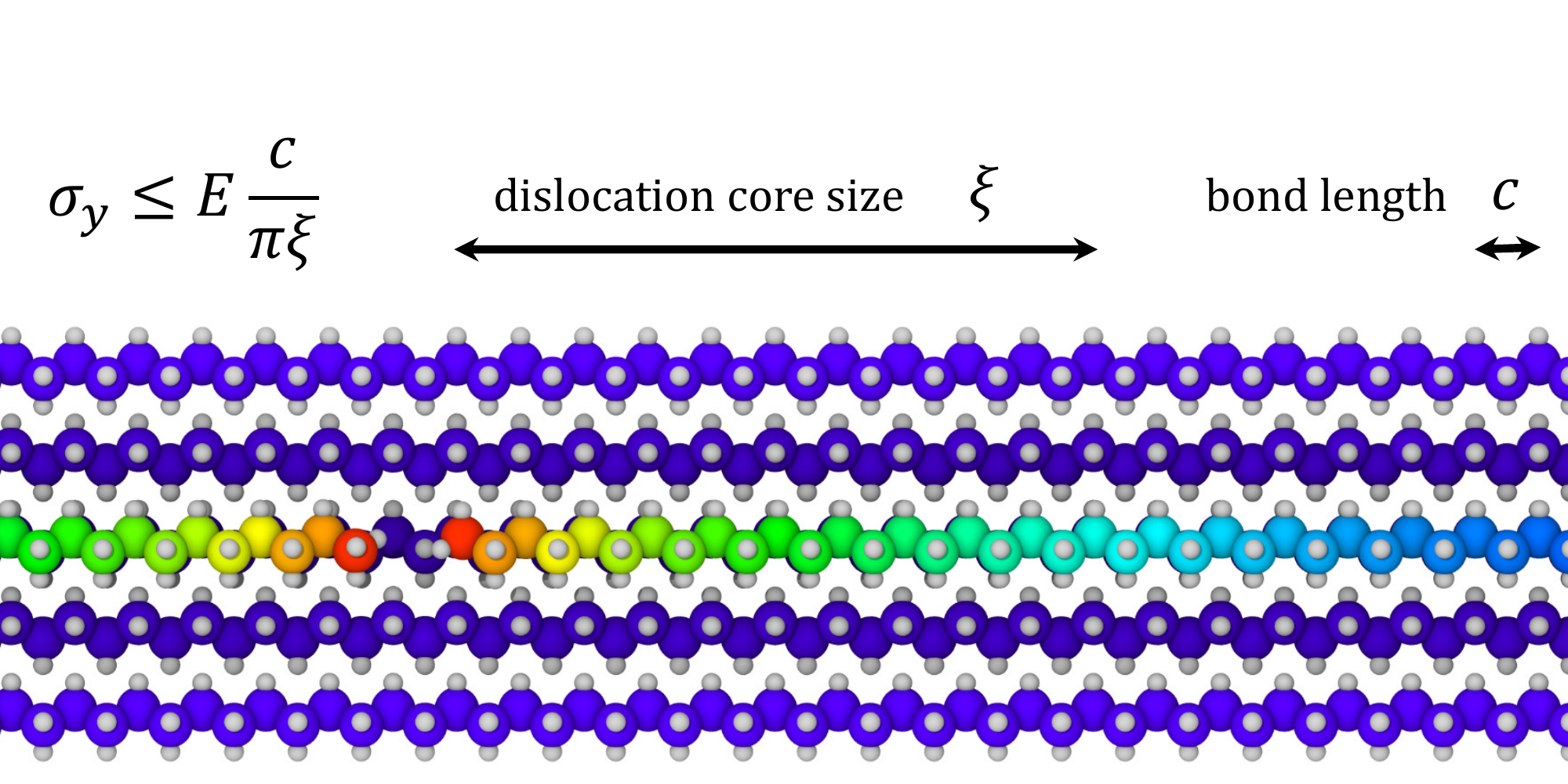}
\label{TOC}
\end{figure}

\end{document}